# A big-data spatial, temporal and network analysis of *bovine tuberculosis* between wildlife (badgers) and cattle

Aristides Moustakas[1,*] & Matthew R Evans[1]

1. School of Biological and Chemical Sciences

Queen Mary University of London

Mile End Road, E1 4NS, London, UK

*Corresponding author:

Aris Moustakas

arismoustakas@gmail.com

**Abstract**

Bovine tuberculosis (TB) poses a serious threat for agricultural industry in several countries, it involves potential interactions between wildlife and cattle and creates societal problems in terms of human-wildlife conflict. This study addresses connectedness network analysis, the spatial, and temporal dynamics of TB between cattle in farms and the European badger (*Meles meles*) using a large dataset generated by a calibrated agent based model. Results showed that infected network connectedness was lower in badgers than in cattle. The contribution of an infected individual to the mean distance of disease spread over time was considerably lower for badger than cattle; badgers mainly spread the disease locally while cattle infected both locally and across longer distances. The majority of badger-induced infections occurred when individual badgers leave their home sett, and this was positively correlated with badger population growth rates. Point pattern analysis indicated aggregation in the spatial pattern of TB prevalence in badger setts across all scales. The spatial distribution of farms that were not TB free was aggregated at different scales than the spatial distribution of infected badgers and became random at larger scales. The spatial cross correlation between infected badger setts and infected farms revealed that generally infected setts and farms do not coexist except at few scales. Temporal autocorrelation detected a two year infection cycle for badgers, while there was both within the year and longer cycles for infected cattle. Temporal cross correlation indicated that infection cycles in badgers and cattle are negatively correlated. The implications of these results for understanding the dynamics of the disease are discussed.

**Keywords**

Network connectedness; point pattern analysis; individual based models; veterinary epidemiology; spatial analysis; temporal correlation

**Introduction**

The control of bovine tuberculosis (TB) in cattle is clearly controversial in Great Britain, with the recent introduction of pilot badger culls raising the intensity of political debate. While the opposing sides argue for tighter controls on either cattle or badgers, the science is often lost. Interpreting scientific uncertainty in this political landscape is challenging. The financial and welfare costs of TB are substantial and have risen along with the incidence and geographical spread of the disease in the cattle herd since the 1970s. Bovine TB has serious financial and welfare costs, in 2013-14 costs were over £99 million, excluding policy development costs (DEFRA, 2014). There are also additional financial and other costs for the farming community and the society in general (Enticott, 2001) despite the fact that it rarely poses a threat to humans (Torgerson & Torgerson, 2008). Efforts have been made since the Second World War to reduce the impact of the disease, while these were initially successful and the disease reached a minimum in the 1970s it has been increasing in both incidence and prevalence since then (Abernethy *et al.*, 2013; Moustakas & Evans, 2016).

It is clearly in the country's interests to reduce the incidence and prevalence of the disease yet these are both steadily rising. Many of the stakeholders in this problem have polarised views and as such the problem has political dimensions that are similar to those around the debate about climatic change (Caplan, 2012; Nisbet & Markowitz, 2015). Stakeholders are broadly divided into those that feel that a wild animal reservoir host (in GB this would be badgers) is the principal cause of disease outbreaks by reintroducing infection to otherwise clean cattle herds, and those who feel that it is to cattle that attention needs to be directed and thus there is a conflict (King *et al.*, 2007; Brooks-Pollock *et al.*, 2014; Macdonald & Feber, 2015) – see also farmers view (Cowie *et al.*, 2015; Enticott *et al.*, 2015; O'Hagan *et al.*, in press). The role of science in this debate should be to provide objective advice and guidance on the best way to tackle the disease. There have been attempts to provide scientific input to this debate; TB was the subject of a large scale ecological experiment – the randomised badger culling trial (RBCT) that ran from 1998 to 2008 (Bourne *et al.*, 2007). Unfortunately the results of this experiment have been interpreted as both providing evidence for and against the utility of badger culling as a means of TB control (Donnelly *et al.*, 2006; King *et al.*, 2007; Donnelly *et al.*, 2015; Donnelly & Woodroffe, 2015). Bovine TB is a problem in several other countries, notably New Zealand (Tweddle & Livingstone, 1994), some states of the USA (Schmitt *et al.*, 1997), and Ireland (Byrne *et al.*, 2015) where there are also tensions between the degree of importance of wild reservoir hosts and cattle (Krebs *et al.*, 1998; Macdonald & Feber, 2015).

Understanding the spread of any disease is a highly complex and interdisciplinary exercise as biological, social, geographic, economic, and medical factors influence the way a disease moves through a population and options for its eventual control or eradication (Christakos *et al.*, 2006; Moustakas & Evans, 2016). Space-time epidemiology is based on the concept that various characteristics of the pathogenic agents and the environment interact and alter the probability of disease occurrence and result in the formation of temporal or spatial patterns (Knox & Bartlett, 1964; Christakos *et al.*, 2014). Patterns of disease occurrence provide insights into which factors may be affecting the health of the population, through investigating which individuals are affected, where those individuals are located and when they become infected (Lange *et al.*, 2014; Alvarez *et al.*, 2016). With the rapid development of computationally-intensive modelling methods (Moustakas & Evans, 2015), smart sensors (Reis *et al.*, 2015), social networks (Weber *et al.*, 2013), digital maps and remotely-sensed imagery (Touloudi *et al.*, 2015), as well as inventory datasets (Schumm *et al.*, 2015) spatio-temporal data are more ubiquitous and richer than ever before in veterinary and ecological epidemiology (Pfeiffer & Stevens, 2015). The availability of such large datasets (big data) poses great challenges in data analysis (Fan *et al.*, 2014; Najafabadi *et al.*, 2015) but may provide insights into the TB problem in the UK in terms of facilitating understanding of the dynamics of the disease and potential approaches to control.

In the current study we sought to quantify the dynamics of TB between and within and between badger and cattle populations using outputs from a stochastic, spatially-explicit model (Moustakas & Evans, 2015). The model has been calibrated in detail with a scale-specific parameter space corresponding to the surface area of a large county in GB, and agent based models of badgers are coupled with agent based models of cattle and farms. We sought to investigate: (i) how connected TB infected badger setts are with other TB infected badger setts, and similarly how connected TB infected cattle farms are with other TB infected cattle farms. (ii) The mean contribution of an infected badger or cattle individual in the spread of the disease in space per annum. (iii) The spatial distribution of TB infected badger setts and TB infected farms, and the cross correlation of TB infected setts and farms across scales. (iv) The temporal correlation of the number of infected badger individuals, the temporal correlation of the number of cattle individuals and their temporal cross correlation.

**Methods**

*Model description*

The model (Moustakas & Evans, 2015) employs agent based models (ABM); (Moustakas & Evans, 2013; Zhang *et al.*, 2015; Augustijn *et al.*, in press) that are coupled (Louca *et al.*, 2015) across scales (Walpole *et al.*, 2013). The model includes both badgers and cattle, it has details of the behaviour and natural history of badgers and the husbandry of cattle and dynamics of TB between and within species - see (Moustakas & Evans, 2015) for a full description. In brief; the model couples an ABM of cattle and farms with an ABM of badgers on a scale-specific grid for each life form. No barriers exist for badgers (they can move throughout the grid), while cattle can move within the farm cells and between farms when a cattle is sold. The grid size of badgers scales to the mean home range area of badgers in the UK (0.7 km$^2$) and the grid size of cattle scales to the farm surface areas in the UK (0.7 - 2.8 km$^2$). Badgers live in social groups (sett size 2-21 individuals), occupying their home range area, they breed in the spring producing an average of three offspring, they live for an average of five years when healthy and up to three years when infected and contract TB both from each other and from cattle (badger-to-badger infection rate = 0.01 - 5%, most common value = 2.55%, badger-to-cattle infection rate = 0.1 - 6%, most common value = 3.4%). Cattle live on farms, give birth of one calf once a year, all the cattle on a farm are housed in one group (one cell of 0.7 km$^2$) over winter but spread around the farm in non-winter months (winter months = November to April), they live on average five years when healthy and up to two years when infected, cattle contract TB from each other and badgers (cattle-to-cattle infection rate = 2.7%, cattle-to-badger infection rate =0.1-6%, most common value = 3.4%). In addition cattle are moved between farms (Gilbert *et al.*, 2005) at realistic rates and distances (percentage of the total cattle population moving to other farms 7.5 – 12% yr$^{-1}$, mean distances of cattle movement 25.2 – 126 km yr$^{-1}$). The TB test in cattle is imperfect i.e. it exhibits false negative cases as it is known to do (Claridge *et al.*, 2012). Cattle are tested for TB prior to movement and at regular intervals (testing every 1 or 4 yr, testing accuracy 50 – 90% most common value 70%). If a cattle individual tests positive it is killed. Badgers can be culled (in 169km$^2$ blocks) and show post-cull perturbation (Woodroffe *et al.*, 2006a) moving up to 3.36 km away from their sett per time step. The time step is one month. The model explicitly follows all individuals. This model is run on a grid of 128 x 128 cells equivalent to the surface area of a large county in the UK, with each cell scaling at 0.7km$^2$, the mean home range area of a group of badgers. The duration of each simulation run is 360 months. The model is initialised with an initial number of badgers and cattle scaling to the mean badger density per unit area (cell) and the mean cattle density per unit area in a county in the UK. Initialization is not homogeneous and it is defined by the initial percentage of infected individuals for badgers or cattle, as well as the min and max group size for badgers. Each simulation scenario follows for each month for 30 years 65,372 badger individuals and 1,387,725 cattle individuals. These numbers derive from a scale-specific model calibration such as there are 5.7 badgers km$^{-2}$ (a value corresponding to Bristol area see table 3.3 in (Krebs *et al.*, 1997)) x 0.7 km$^2$ cell$^{-1}$ (this is the cell size in the model) x 128 x 128 cells (the simulation grid) and for cattle (1,387,725 = 121 cattle km$^{-2}$ (Eurostat, 2009) x 0.7 km$^2$ cell$^{-1}$ x 128 x 128 cells). Full references and a detailed description of how these parameters were derived are provided in Supp. 1 in Moustakas and Evans (2015). The amendments in the parameters space from Moustakas and Evans (2015) performed here due to additional data availability as well as the number of sufficient simulation scenarios n=161, determined from Latin Hypercube Sampling are described in the next paragraph. Each of the 161 simulation scenarios was replicated 10 times to account for model stochasticity. We have thus followed in a spatially-explicit manner a very large number of badgers and cattle (65,372 badgers and 1,387,725 cattle for 360 months across 161 scenarios, each scenario replicated 10 times).

Previous model outputs have been compared with known patterns of TB see section 'Confronting model outputs with data' in (Moustakas & Evans, 2015).

*Model parameterisation*

We used a large and computationally intensive parameter space as described in (Moustakas & Evans, 2015) and supplementary material therein . The model's sensitivity to input parameters had been tested with Latin Hypercube Sampling a robust method for sweeping out parameter space (McKay *et al.*, 1979; Athira & Sudheer, 2015). Two amendments in the input parameters of (Moustakas & Evans, 2015) were made here: (i) the initial number of infected badgers, and (ii) initial number of infected cattle. Recent papers provide field evidence that there are locations in the UK where the percentage of infected badgers is as high as 50% or even higher (Weber *et al.*, 2013; King *et al.*, 2015). We have expanded parameter space by including scenarios in which the percentage of initially infected badgers is up to 55% (initial percentage of infected badgers 2 – 55% most common value 4.05%). Public data show that the percentage of cattle herds that are not TB free has been detected to be as high as 5.4% in July 2014 in the UK and in high risk areas up to 14.5% (DEFRA, 2015). We have expanded parameter space by including scenarios with percentage of initially infected cattle up to 15% (initial percentage of infected cattle 0.2 – 15%, most common value = 2%).

*Analysis*

*Infection networks*

We sought to quantify how connected infected badgers are with other infected badgers and how connected infected cattle are with other infected cattle. To do so Krackhardt's connectedness was calculated for a non-directed graph *G* (Krackhardt, 1994). Connectedness is equal to the fraction of all pairs, (*i, j*), such that there exists an undirected path from *i* to *j* in *G*. Values of connectedness close to 1 indicate that all nodes (in our case badger setts or farms) are connected with every other node in the graph while values close to 0, indicate that the nodes in the graph are isolated. Analysis was conducted using the 'connectedness' function in the 'sna' package in R (R Development Core Team, 2016). Connectedness was calculated for networks of badgers and cattle separately at the final time step of each simulation scenario. For badgers a node in the graph was a badger sett containing at least one infected badger. For cattle a node in the graph was a farm containing at least one infected cow.

*Distance of infections in time*

We quantified the mean distance between an infected individual and any other individual(s) it infects per year. Each individual that was either infected at the beginning of the simulation or that became infected at some time step during the simulation was followed. We recorded the starting cell (*i. j*) of each initially infected individual, or the cell in which it became infected (each then became a focal individual). For each time step (month) the uninfected individuals that were infected by the infected focal individual and the coordinates of the cell in which they contracted the infection were recorded. At the end of the simulation period we summed the distances between the initial location of the focal individual (either the starting point if the focal individual was infected from the beginning of simulation or the cell in which the individual became infected) and the locations of all the individuals which contracted infection from the focal individual (where they first became infected). For each focal individual we divided the sum of these distances in terms of number of cells by 30 years equal to the length of each simulation scenario, and multiplied by the scale of cell side (= 0.84 km) resulting in a mean distance of infection in km yr$^{-1}$. This process was repeated for both cattle and badgers across all scenarios, and the mean of these values was calculated for each scenario.

*Spatial point pattern analysis*

In order to quantify the finer scale spatial pattern of the disease we calculated the pair-correlation function *g(r)*, (Stoyan & Stoyan, 1994). The function *g(r)* is the conditional probability density of finding an infected individual at a distance *r* , given that there is an infected individual at the coordinate origin. The function *g(r)* provides a measure of local spatial ordering which is the mean expected density *V* of points *N* at a given distance *r* from an arbitrary point $r_{ij}$, divided by the intensity of the pattern. It is defined as:

$$g(r) = \frac{V}{4\pi r^2 N^2} [\sum_i \sum_{j \neq i} \delta(r - r_{ij})]$$

Under complete spatial randomness (CSR), *g(r)=1*. Values of *g(r)< 1* indicate regularity at scale r (also called overdispersion), while values of *g(r)>1* indicate clustering at scale r (also called aggregation). The *g(r)* is non-cumulative and particularly suitable to reveal critical scales of the pattern (Wiegand & Moloney, 2004). In addition we calculated the L-function, a cumulative counterpart to g(r) (Besag, 1977), to assess spatial pattern of the disease at larger scales *r* (Wiegand & Moloney, 2004). *L(r)* is a summary statistic based on Ripley's *K*-function:

$$K(r) = 2\pi \int_{r=0} g(r) r dr \text{ and } L(r) = r(\sqrt{\frac{K(r)}{\pi}} - 1).$$

Under CSR, *L(r)* = 0 and values of *L(r)* < 0 indicate regularity, while values of *L(r)* > 0 indicate aggregation. Monte Carlo permutations (n=99) were employed to approximate confidence intervals of the *L(r)*.

The *g(r)*, and *L(r)* analysis was performed for the final time step of each simulation scenario for badgers and cattle separately, and the distance-specific *g(r)* and *L(r)* values were averaged across scenarios and plotted - for a similar approach see (Moustakas, 2015). For badgers the locations of setts containing at least one infected badger were analysed, while for cattle the locations of farms containing at least one infected cow were analysed. Analysis was conducted using the 'pcf' and 'Lest' functions respectively in 'spatstat' library in R (R Development Core Team, 2016). We also sought to quantify the scale of spatial coexistence of TB infected badgers and cattle. We analysed the spatial cross-correlation between locations (points) of badger setts containing at least one infected badger and locations of farms containing at least one infected cattle using a multivariate spline cross-correlogram (BjØrnstad & Falck, 2001). Spatial cross-correlogram estimates the spatial dependence at discrete distance classes. The region-wide similarity forms the reference line (the zero-line); the x-intercept is thus the distance at which object are no more similar than that expected by-chance-alone across the region. The analysis was performed using the 'spline.correlog' function in the 'ncf' package in R (R Development Core Team, 2016).

*Temporal analysis*

We sought to quantify the temporal dynamics of TB in badgers, cattle, and between them. To do so we calculated the partial correlation function for infected badgers and infected cattle and the temporal cross correlation between infected badgers and cattle (Venables & Ripley, 2002). Temporal correlation analysis was performed using the time series of the number of infected badgers and infected cattle on a monthly time step throughout each simulation. All data were $\log_{10}(x+1)$ transformed prior to the analysis to successfully normalise them. The analysis was performed using the 'acf', 'pacf', and 'ccf' functions in the 'stats' package in R (R Development Core Team, 2016). As we had multiple observations for each month (equal to the number of simulation scenarios explored), we averaged the values for each time step.

**Results**

*Infection networks*

Badger infection networks had mean connectedness values of 0.12 with a standard deviation of 0.09 (Table 1). Cattle networks had mean connectedness values of 0.23 with a standard deviation of 0.12 (Table 1). Cattle infection networks are therefore relatively more interconnected than badger infection networks.

*Distance of infections in time*

The mean distance that an infected badger individual spreads TB is 0.92 km year$^{-1}$, StDev = 0.62 (Fig. 1a). The mean distance that an infected cattle individual contributes into the spread of the disease is 2.34 km year$^{-1}$, StDev = 0.98 (Fig. 1a). The distribution of cattle individuals' distances of disease spread varied greatly; the distribution peaks at three values (ca 0.75km, 1.5km and 2.25 km) and has a long tail of long distance disease spread values (Fig. 1a). The cumulative effect of the differences between the distribution of mean infection distances can be seen in figure 1b which displays the empirical cumulative density function (ECDF) and shows that 80% of badgers spread the disease at distances up to one km per year (Fig. 1b). Around 80% of cattle spread the disease at distances up to three km per year (Fig. 1b).

Further analysis of the data reveals that TB spread in badgers derived mainly from badgers leaving their sett: a contour plot of badgers leaving their setts vs badgers remaining in their sett and the distance that badgers spread TB per year showed that around 17,000 moving badgers contribute to the same mean distance of disease spread as 225,000 non moving badgers (Fig. 1c). The number of infected badgers is positively correlated with the badger population growth rate (SS=4065120000, F=26,12, p<0.001, $R^2$=13.6%; Fig. 1d). The number of infected badgers is positively correlated with the number of badgers leaving their setts (SS=5797950000, F=40.06, p<0.001, $R^2$=19.6%; Fig. 1e), more badgers leave their home sett as badger population growth rate increases (SS= 1084865345, F= 69.74, p<0.001; $R^2$= 30.1%, Fig. 1f) and the distance that badgers spread TB increases as badger population growth rates increase (Fig. 1g). However, the percentage (fraction) of infected badgers was not significantly correlated with badger growth rate (SS=0.2378, F=3.45, p=0.065, $R^2$=1.5%; Fig. 1h), was negatively correlated with the distance that badgers spread TB in a year (SS=0.201758, F=78.05 p<0.001, $R^2$=32.5%; Fig. 1i) and negatively correlated with the number of badgers leaving their home sett (SS= 796055533, F=45.82, p<0.001, $R^2$=21.9%; Fig. 1j).

The distance that cattle spread TB per year increased as the percentage of cattle individuals moved from their current farm rises (Fig. 1k).

*Spatial point pattern analysis*

An overview of the spatial distribution of badger setts with at least one infected badger and of cattle farms with at least one infected cattle in Fig. 2a. A detail of spatial TB dynamics with a Gaussian kernel smoothing with a sigma equal to the standard deviation to improve the visualization of setts with at least one TB infected badger is provided in Fig. 2b, of farms with at least one TB infected cattle in Fig. 2c. These provide a visualisation of the differences in the spatial patterns of the disease in the two species after 30 years of simulation.

The spatial distribution of setts containing at least one TB infected badger was aggregated at scales up to 4.3 km (Fig. 3a), and cumulatively aggregated across all examined scales up to 25.2 km (Fig. 3b). The spatial distribution of farms containing at least one TB infected cattle was aggregated at scales up to 11.5 km (Fig. 3c), and cumulatively aggregated at scales up to 21.1 km and random thereafter (Fig. 3d). The spatial distribution of TB in both species (at least one infected badger per sett or at least one infected cattle per farm) was aggregated at scales up to 4.4 km (Fig. 3e), and cumulatively aggregated across all scales examined up to 25.2 km (Fig. 3f). This shows that infected badger setts are typically clustered together, while infected cattle farms are clustered at small scales but become randomly distributed throughout space at large scales.

Results from spatial cross-correlation indicated that overall there are few scales where infected badgers and infected cattle coexist; the cross correlation between badgers and cattle was weakly positive and not significantly different from zero at distances up to 6.3 km (the lower confidence interval crosses zero) (Fig. 3g). Cross correlation between badgers and cattle was positive and significant at distances between 7.5 km and 9.25 km (Fig. 3g), and negative or non-significantly different from zero at all other scales (Fig. 3g).

*Temporal analysis*

Temporal autocorrelation was significantly positive for badgers for time lags of up to 23 months (Fig. 4a). Temporal autocorrelation in cattle was significantly positive for time lags of one to 7 months, and significantly positive for time lags of 32 to 39 months (Fig. 4b). Temporal partial autocorrelation showed no significant correlation cycles for the number of infected badgers (Fig. 4c), but significantly positive cycles for the number of infected cattle for time lags of 31 to 36 months, and a significant negative cycle for a time lag of 37 months (Fig. 4d). Temporal cross correlation between the number of infected badgers and cattle was always negative across all time lags, with this negative effect being more pronounced between lags of 1 to 6 (Fig. 4e).

**Discussion**

Computational models provide a valid alternative to limitations due to accessibility, ethics, or cost-prohibitive experimental approaches (Desouza & Yuan, 2013) and present a method of testing the likely effects of various strategies designed to control or eradicate TB in cattle (Brooks-Pollock *et al.*, 2014; Moustakas & Evans, 2015). To be useful such models need to represent the modelled system and be parameterised in sufficient detail (Evans *et al.*, 2013; Evans *et al.*, 2014; Lonergan, 2014) to allow realistic predictions to be made about the outcome of any control strategy.

*Badger growth rates and TB spread*
It is well recorded that badger culling induces perturbation thereby increasing badger dispersal (Woodroffe *et al.*, 2006a; Woodroffe *et al.*, 2006b; Carter *et al.*, 2007; Donnelly *et al.*, 2007; Jenkins *et al.*, 2010). Several studies have shown that cattle testing is an imperfect but effective control strategy (Goodchild *et al.*, 2015), that cattle testing is more effective than badger culling as a control strategy (Brooks-Pollock *et al.*, 2014; Moustakas & Evans, 2015) and that every control strategy should be implemented such as to minimise badger population disruption (Wright *et al.*, 2015). It has been reported that group size reduction rather than group size *per se* has most influence on disease dynamics (Vicente *et al.*, 2007). Results derived here show that badger population growth is correlated with the number of badgers leaving their home sett and that they then contribute more to the spread of the disease than non-moving badgers, despite the number of non-moving badgers being at least an order of magnitude higher than moving badgers. To that end high population growth rates in badgers may act similarly to culling: a high number of dispersing badger individuals spread the disease more than a low number of dispersing badger individuals, and the number of dispersing individuals increases both with culling and with a high badger population as both result in badgers leaving their home sett. While this applies to the number of badger individuals, it is not valid for percentages: there was no significant correlation with badger growth rates and the percentage of infected badgers, and there was negative correlation between the percentage of infected badgers and the absolute number of badgers leaving their home sett. This study does not address why badgers are increasing in some regions. Badger density in the Republic of Ireland is not increasing (Byrne *et al.*, 2012; Byrne *et al.*, 2014a) nor in Wales (Judge *et al.*, 2014), but it is increasing in England (Judge *et al.*, 2014). Badger densities in England are already high (some recently reported values span from 5.2 badgers km$^{-2}$ (Smith & Cheeseman, 2007) to 11.6 badgers km$^{-2}$ (Woodroffe *et al.*, 2008)) in comparison to Ireland (1-3 badgers km$^{-2}$ (Byrne *et al.*, 2012; Byrne *et al.*, 2014a)). Badgers are among the top predators in the habitats in which they live, but they are generalists with a wide diet range; despite this their population size will be limited by food availability. Therefore it would be logical to couple badger population changes over time with data on land use changes over time, to discern any possible causal patterns which might explain this population growth.

*Networks of infections and the contribution of individual badger and cattle to TB spread*

According to results derived here, cattle farms with at least one infected individual are more strongly connected with each other than badger setts containing at least one infected badger are connected with other infected setts. Furthermore, the contribution of infected badger individuals to the spread of the disease in terms of km yr$^{-1}$ is considerably smaller than the contribution of infected cattle. Values of connectedness between farms derived here are lower than the ones derived for farms in Minnesota, USA (Ribeiro-Lima *et al.*, 2015) possibly due to lower cattle movements in Minnesota. Studies of networks of badgers have reported that badger networks correlate with infections (Weber *et al.*, 2013). A study of TB in possums in New Zealand reported that potential contact with TB-positive possums increased the odds of disease transmission whereas potential contact with a large number of possums did not (Porphyre *et al.*, 2011). This suggests that multiple contacts with TB infected possums are necessary for transmission of TB and this is more likely to occur in networks that are smaller (Porphyre *et al.*, 2011). While there are obviously differences between possums and badgers this is likely to have some merit also in badger networks in the UK, as it is consistent with the finding that TB prevalence was consistently higher at low badger densities and in small social groups (Woodroffe *et al.*, 2009).

Analysis of cattle movements in the UK has indicated that this movement is contributing towards the spread of the disease (Gilbert *et al.*, 2005; Gopal *et al.*, 2006) and that the majority of cattle movements occur over a range of 10 to 100 km per journey (Christley *et al.*, 2005) although many tens of thousands of cows move over far greater distances (up to 1000 km) (Mitchell *et al.*, 2005). The geographical distribution of these movements appears to be relatively stable from year to year (Mitchell *et al.*, 2005). The distribution of annual rate of disease spread by cattle in the model is close to the values obtained in another study for the spread of TB in cattle (0.04 - 15.9, median = 3.3 km per year), and the same study additionally found a long tail (Brunton *et al.*, 2015).

Analysis of badger movement patterns has indicated that generally badgers move very little outside their home range area: Inter-group contacts between badgers only occurred between directly adjacent social groups at a frequency <1% of all contacts, in a medium density population (O'Mahony, 2015), and even in a high density population the majority (75.8%) of badgers were never captured in more than two social groups, (Macdonald *et al.*, 2008). Despite the existence and potential underestimation of long-distance badger dispersal, in a field study addressing badger long-distance dispersal, the longest recorded was 22.1 km with a mean distance of 2.6 km and with the 95 percentile at 7.3 km (Byrne *et al.*, 2014b). These dispersal values are considerably smaller than the distance that cattle disperse (Christley *et al.*, 2005; Gilbert *et al.*, 2005; Mitchell *et al.*, 2005). In addition contacts between badgers and cattle occurred more frequently than contacts between different badger groups and these inter-specific contacts involved those individual cows, that were highly connected within the cattle herd (Böhm *et al.*, 2009). These differences in the distances dispersed by the two species involved here are consistent with the differences in the patterns of TB spread by the two species. Badgers disperse short distances, typically interact only within social groups and spread TB relatively short distances. Cattle are moved large distances, interact within the herd in which they reside and spread TB over both long and short distances.

*Spatial pattern analysis*

Results derived here show that there is a scale-dependent pattern in the spatial distribution of farms that contain TB infected cattle. TB infected farms aggregated at scales of 11.5 km corresponding to infections between nearby farms and at larger scales up to 21.1 km but were random thereafter. Movements of cattle from farm to farm via markets should also be considered at such distances (see also Fig. 11 in Mitchell et al. 2005) however this model does not include that. The spatial distribution of setts containing TB infected badgers exhibited an aggregated pattern across all scales examined with local infections recorded at scales of up to 4.3 km. The spatial distribution of TB (including both setts and farms) was aggregated at all scales with local infections recorded at scales up to 4.4 km. Therefore the TB spatial prevalence pattern is different between infected farms and setts but the overall TB infection spatial pattern is similar to the one of the most aggregated pattern – badgers. This implies that spatial associations between infected badgers and cattle will be detected at fine scales of 1-2 km (Woodroffe *et al.*, 2005; Byrne *et al.*, 2015), but locations of infected cattle will not detect infected badgers (Smith *et al.*, 2015) and *vice versa*. In a study of farms in the UK it was concluded that cattle tested and found infected (reactors) within 8 km of known infected herds are more likely to harbour undetected infection than those located further away (Goodchild *et al.*, 2015) indicating a potential local clustering at a scale comparable to the one derived here for infected farms.

These results have implications for TB vaccinations of badgers (Carter *et al.*, 2012): Spatial pattern formation facilitates eradication of infectious diseases and disease vaccination efforts could be spatially targeted to prioritize those areas where the disease is known to occur (Eisinger & Thulke, 2008). Field evidence showed that vaccination reduces risk of TB infection in badgers (Carter *et al.*, 2012) and surveys showed that some farmers in the UK are becoming more positive towards it (Enticott, 2015; O'Hagan *et al.*, in press) but see also (Naylor *et al.*, 2015). Equally for cattle vaccination, a spatial targeting following the pattern and scale of not TB free farms can save effort in vaccination and towards disease control (Eisinger & Thulke, 2008). Note that until recently cattle vaccination was problematic because it was hard to distinguish infected from vaccinated cattle (Vordermeier *et al.*, 2016). However recently tests have been developed that differentiate infected from vaccinated animals, and it may now be feasible to use vaccines to assist in the control of this disease (Parlane & Buddle, 2015; Vordermeier *et al.*, 2016).

Results derived here indicate generally little overlap between infected setts and infected farms and that this overlap mainly occurs in a few scales; the spatial cross correlation between TB infected setts and infected farms was only positive at scales of 7.5 - 9.25 km and either not different from zero or negative at other scales. Thus there is a fairly narrow total area defined by a torus of width of 1.75 km between infected setts and infected farms where both TB positive cattle and badgers coexist. This is in agreement with field studies reporting direct contacts between badgers and cattle at pasture being very rare (four out of >500 000 recorded animal-to-animal contacts) despite ample opportunity for interactions to occur (Drewe *et al.*, 2013), that badgers avoid cattle (Mullen *et al.*, 2015), and that direct contact between individuals is unlikely to be a major route of TB transmission between species (Drewe *et al.*, 2013; Mullen *et al.*, 2015). The overlap between infected setts and farms at scales 7.5 – 9.25 km but not at scales where the actual infection is taking place, 1 to 2 km, shows that in the locations where both farms and setts are infected the infection between them is a movement-based one rather than a residential one deriving from either cattle or badgers that move. These results indicate that the main source of infection for badgers is other badgers, and for cattle are other cattle. For data analyses on cattle to cattle transmission see (Conlan *et al.*, 2012) while for badger to cattle see (Donnelly & Nouvellet, 2013).The introduction of the disease in areas that are TB free is more likely to occur through cattle movement as infected farm networks are better connected and cattle move considerably larger distances than badgers, and there are considerably more cattle than badgers per unit area.

*Temporal analysis*

Temporal autocorrelation results derived here indicated that there are positive population cycles of infected badgers up to 23 months but no significant cycles in terms of partial autocorrelation indicating no longer term cycles or cycles within the 23-month-cycle. Temporal autocorrelation showed cycles of infected cattle of one to 7 months, of 32 to 39 months, while partial autocorrelation detected cycles 31 to 36 and a significant negative cycle for a time lag of 37 months. Combined these results suggest a (positive or negative) three-year infection cycle - the cycles between 1-7 months can be an artefact created by the three-year cycle. Temporal cross correlation between the number of infected badgers and cattle was always negative across all time lags. These results imply that badgers have infection cycles of around two years and are negatively synchronised with cattle infection cycles as deduced from cross-correlation. Badger life span when infected is less than when not infected, and the time from infection to death is reported to vary between studies spanning from 'a rapid course' up to 709 days (Clifton-Hadley, 1993) and up to 3.5 years when in captivity (Little *et al.*, 1982). This cycle is likely to correspond to a median of two years, which is the value that we have typically used as badger life span when infected (Moustakas & Evans, 2015), and thus the cycle duration may be linked to the average life span of infected badgers. The decay of the autocorrelation function in badgers may in part be caused by the increase of badger population in time in several simulation scenarios. Other studies addressing temporal trends of TB in badgers have reported significant correlation in the disease status within groups over time, suggesting that infection persists for many years in some social groups (Delahay *et al.*, 2000). The same study also reported that temporal trends in disease were not synchronized amongst neighbouring groups, suggesting low rates of disease transfer between them (Delahay *et al.*, 2000). Seasonal trends of TB in badgers were reported to be significantly higher in summer than in any other season, but this was not consistent across all study locations (King *et al.*, 2015).

Cattle have an infection cycle within the year lasting 6 months possibly corresponding to winter housing lasting four months (Moustakas & Evans, 2016) and a longer term oscillation of around three to three and half years. The negative cycle of 37 months in cattle is likely to derive from cattle testing: in the analysis we have averaged several scenarios, these included annual testing when the percentage of infected herds is > 1% and testing every four years (48 months) when infected herds are 0.2% or less according to EU directives. These results indicate that (i) the cycle of infection is different between badgers and cattle, (ii) that the cycle is shorter in badgers than in cattle, and (iii) that there is a negative cycle in cattle infections possibly due to testing, while there is no inverse cycle in badgers.

**Acknowledgements**

Comments of three anonymous reviewers have been exceptionally thorough and helpful. This paper is part of the special issue in Spatio-temporal Data Mining in Ecological and Veterinary Epidemiology. AM has been the guest editor for the special issue and declares that this submission was handled by regular member of the editorial board.

**Table 1**

| Variable | Mean | StDev | Q1 | Median | Q3 |
|---|---|---|---|---|---|
| cattle | 0.23 | 0.12 | 0.14 | 0.22 | 0.32 |
| badgers | 0.12 | 0.09 | 0.04 | 0.11 | 0.17 |

**Table 1.** Connectedness values between farms containing at least one infected cattle (cattle) and between badger setts containing at least one infected badger (badgers).

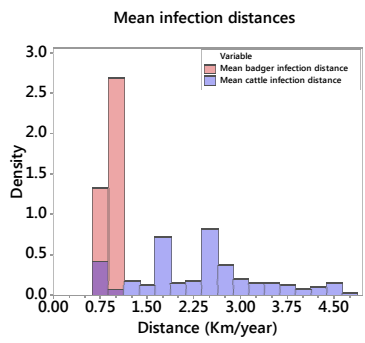

a

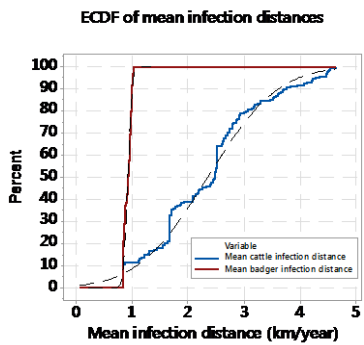

b

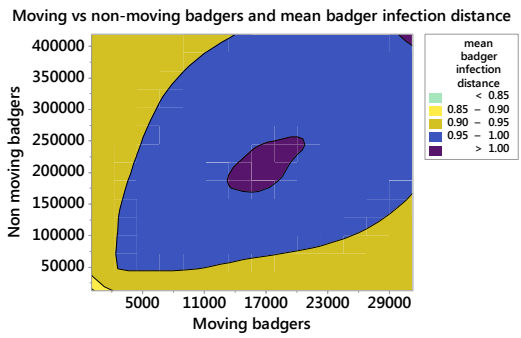

c

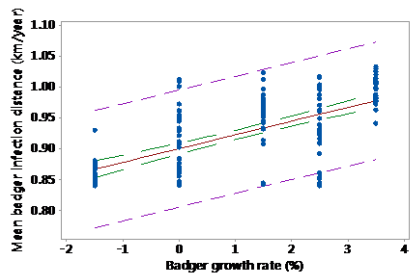

d

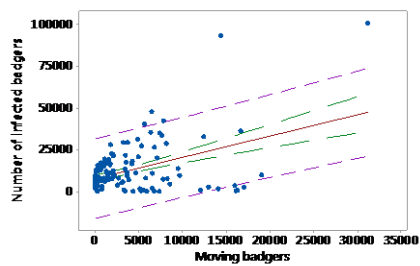

e

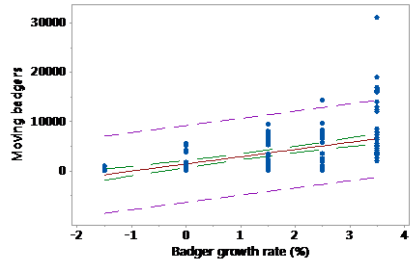
f

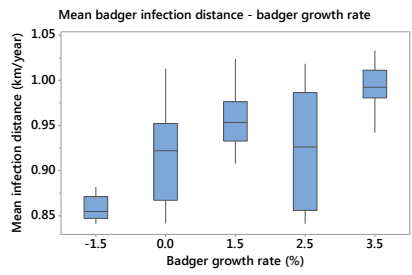
g

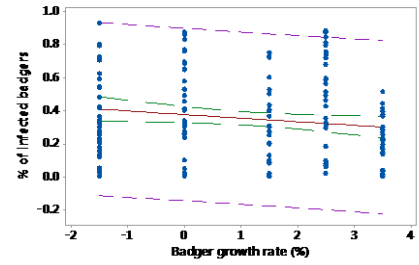
h

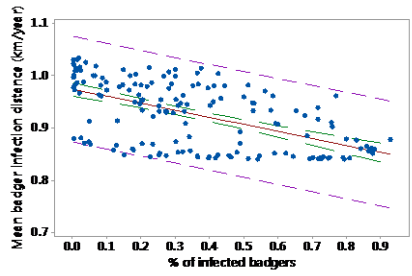
l

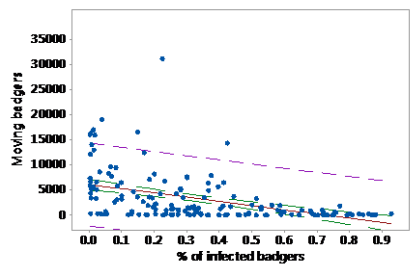
j

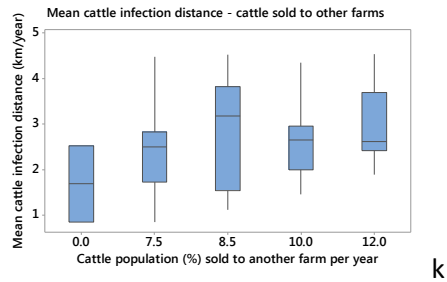

**Figure 1**. Distance of infections in time. (a) Mean distance of infection spread per infected badger and per infected cattle individual per year (km yr$^{-1}$). (b) Empirical Cumulative Density Function (ECDF) of the distribution of mean infection distances for badgers and cattle. ECDF plots the value of each observation against the percentage of values in the sample that are less than or equal to that value. Data are in stepped lines while the continuous line is a fitted normal distribution in the data (c) Contour plots of the contribution of infected badger individuals in terms of mean distance of disease spread (km yr$^{-1}$) of badger individuals staying in their home sett and badger individuals leaving their home sett (moving badgers). Darker colours indicate longer mean distances of disease spread. (d) Linear regression between the infection spread distance (km yr$^{-1}$) and badger population growth rate Regression fit is displayed with solid red lines, 95% confidence intervals are displayed with dashed green lines and 95% predicted intervals are displayed with dotted purple lines (e). Linear regression between the number of infected badgers and the number of badgers leaving their sett (f) Linear regression between the number of badgers leaving their sett and badger population growth rate. (g) Box plots of badger infection distance (km yr$^{-1}$) per badger population growth rates. (h) Linear regression between the percentage (fraction) of infected badgers and badger population growth rate. (i) Linear regression between badger infection distance (km yr$^{-1}$) and the percentage of infected badgers. (j) Linear regression between the number of badgers leaving their sett and the percentage of infected badgers. Note that the lower predicted interval is always below zero but the lower confidence interval is only crossing zero for very large values of % of infected badgers. (k) Box plots of cattle infection distance (km yr$^{-1}$) as a function of the fraction of the total cattle population moving to other farms per year.

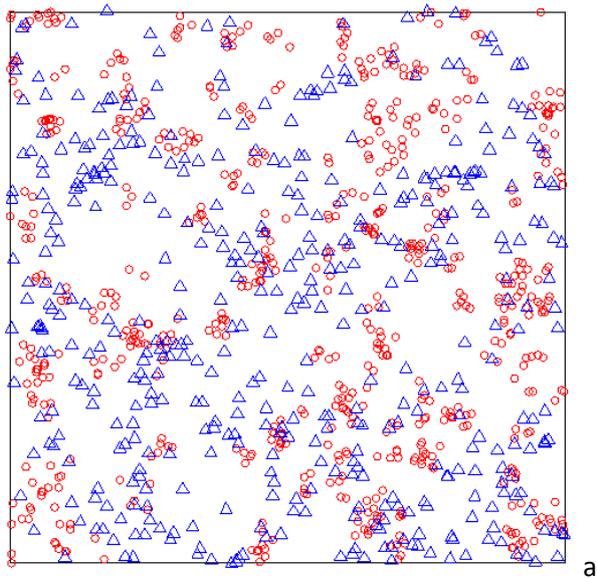

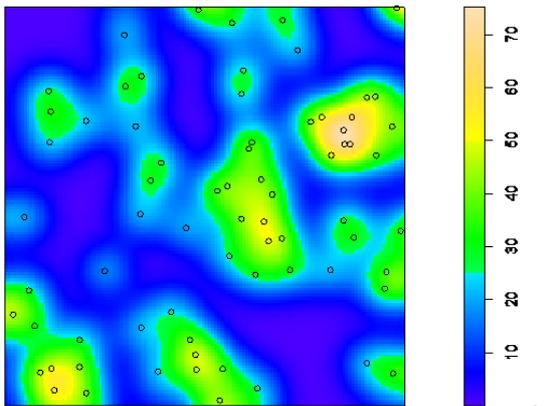

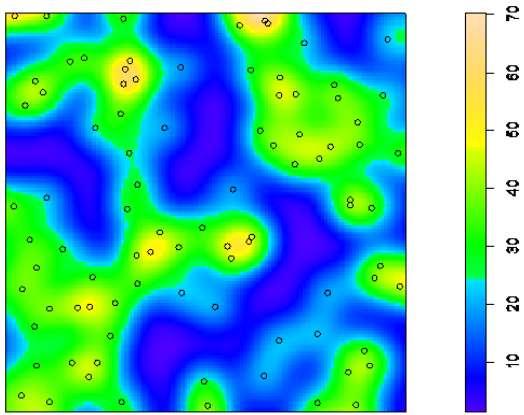

**Figure 2.** (a) Full scale model output of locations of both badger setts and cattle farms with at least one TB infected individual at the end of simulation. Badgers are depicted with red circles while cattle with blue triangles (b) A time snapshot of a detail (20 x 20 cell grid) of locations of badger setts with at least one TB infected badger individual in the middle (month = 180) of simulation period. For improving visualisation, a Gaussian kernel smoothing with a sigma equal to the standard deviation was used. (c) same as b for cattle.

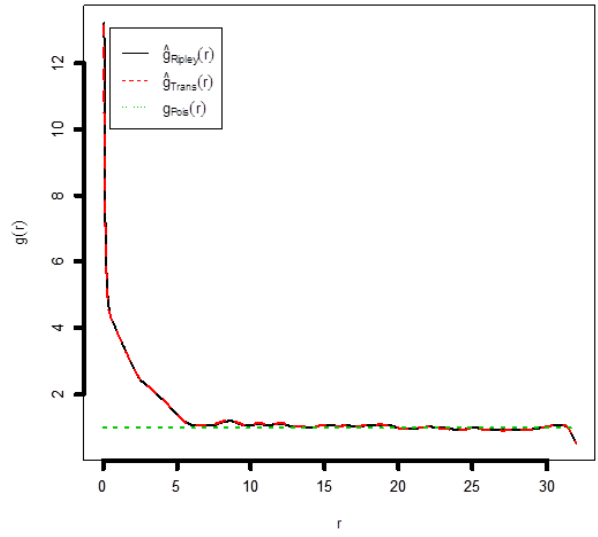

a

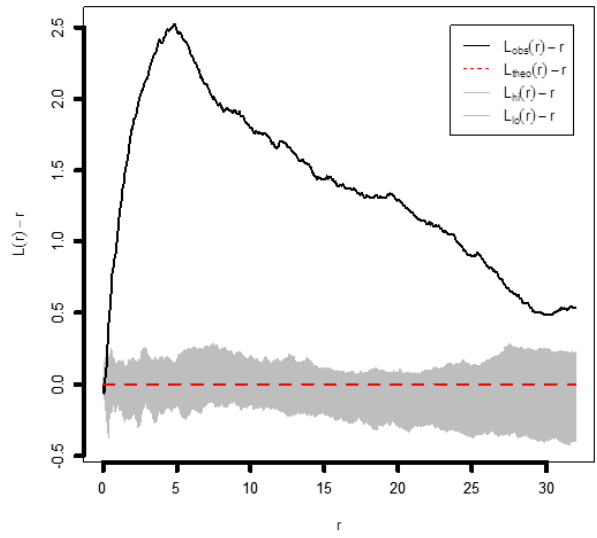

b

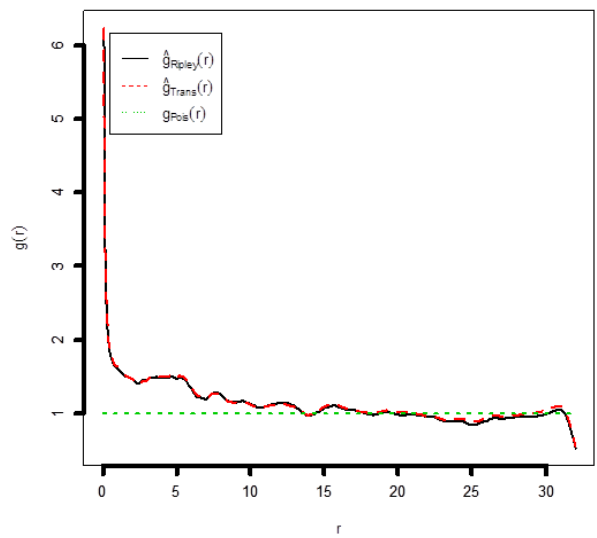

c

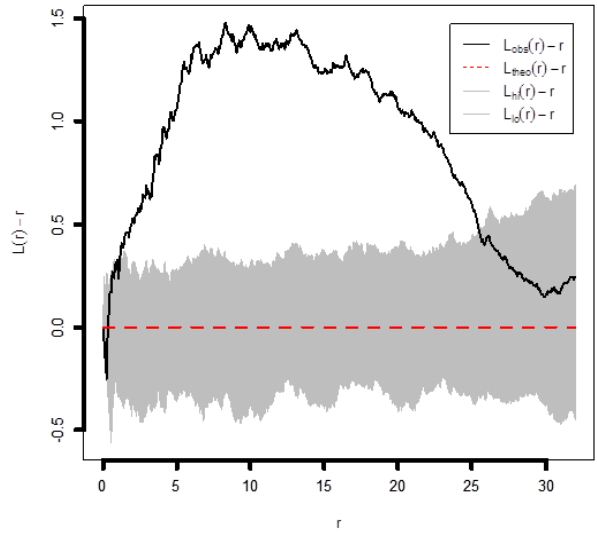
d

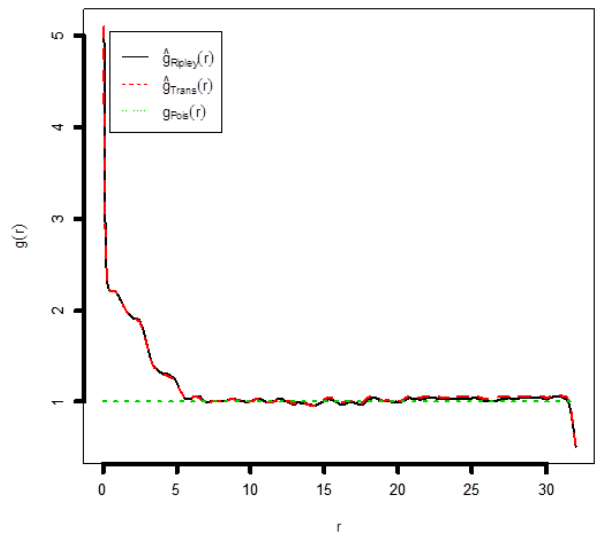
e

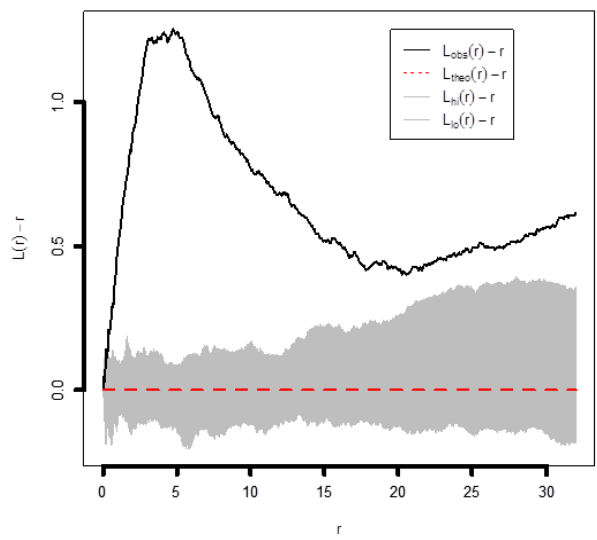
f

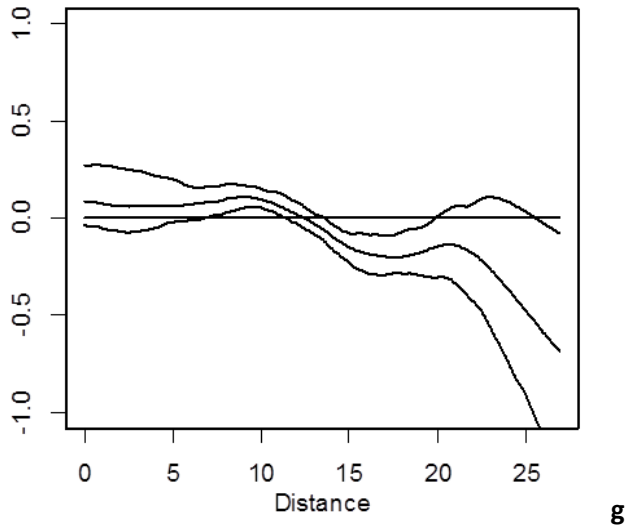

g

**Figure 3**. Spatial analysis results of badger setts containing at least one infected badger and farms containing at least one infected cattle. All horizontal axes refer to distances (scales, *r*) in increments of 0.84 km corresponding to one cell in the model. Upper graphs refer to badgers (a, b), middle graphs to cattle (c, d), and bottom graphs badgers and cattle (e-g). Graphs a, c, and e (first column) refer to the pair correlation function g(r) across scales r. Values of *g(r)< 1* indicate regularity at scale r, while values of *g(r)>1* indicate clustering at scale r, while values close to 1 indicate random spacing. Graphs b, d, and f (middle column) refer to the L-function with confidence intervals in grey colour. Values of the *L(r)*>0 indicate clustering at scale *r* while *L(r)*<0 indicate regularity at scale *r*. Graph g depicts the spatial cross-correlation between locations across scales *r* of badger setts containing at least one infected badger and locations of farms containing at least one infected cattle using a multivariate spline cross-correlogram. Values on the vertical axis spanning from -1 to 1 indicate negative (-) positive (+) or no correlation (≈0). Upper and lower lines depict a 95% confidence interval. In order to assume a scale-specific spatial correlation significance all three lines should be above or below zero (no line should cross zero at that scale).

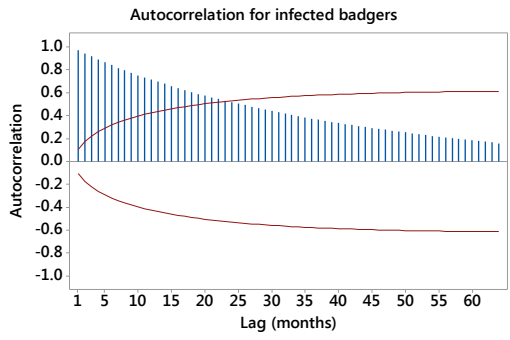

a

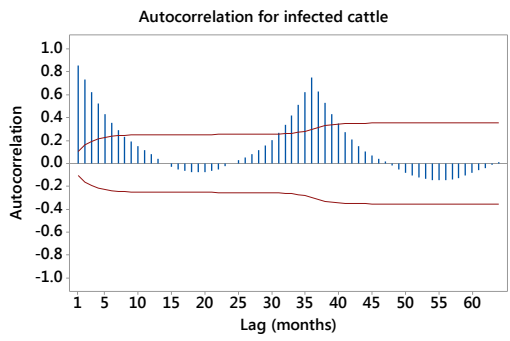

b

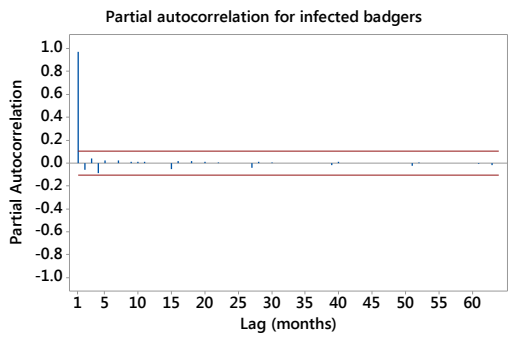

c

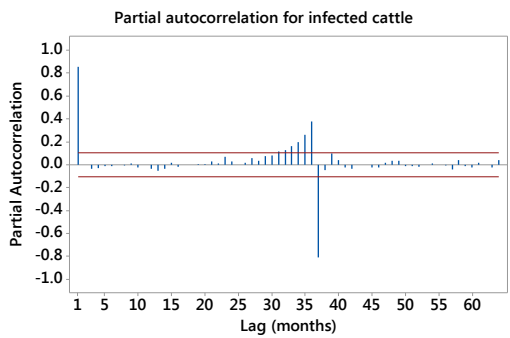

d

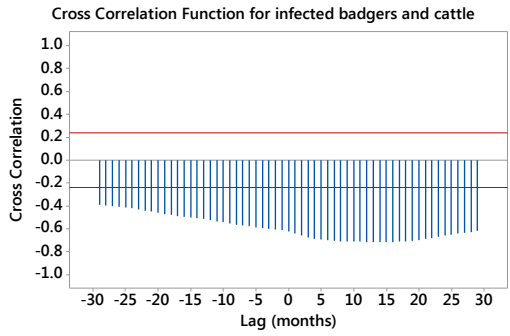

e

**Figure 4.** Temporal analysis of the number of infected badger and cattle individuals per time month for 30 years of simulation. (a) Autocorrelation function for the number of infected badgers per time step of simulations. (b) Autocorrelation function for the number of infected cattle per time step of simulations. (c) Partial autocorrelation function for the number of infected badgers per time step of simulations. (d) Partial autocorrelation function for the number of infected cattle per time step of simulations. (e) Cross correlation function between the number of infected badgers and the number of infected cattle per time step of simulations. Vertical solid red lines indicate 95% confidence intervals.